\documentclass[aps,pre,twocolumn,superscriptaddress,showkeys]{revtex4-2}

\usepackage{amsmath,amssymb,bm}
\usepackage{graphicx,color}
\graphicspath{{/mnt/data/}{./}}
\usepackage[hidelinks]{hyperref}

\newcommand{\R}{\mathbb{R}}
\newcommand{\Hh}{\mathbb{H}}
\newcommand{\dd}{\mathrm{d}}

\begin{document}

\title{Sampling biases in sprinkled causal sets}

\author{Mari\'an Bogu\~n\'a}
\email{marian.boguna@ub.edu}
\affiliation{Departament de F\'isica de la Mat\`eria Condensada, Universitat de Barcelona, Mart\'i i Franqu\`es 1, E-08028 Barcelona, Spain}
\affiliation{Universitat de Barcelona Institute of Complex Systems (UBICS), Universitat de Barcelona, Barcelona, Spain}

\author{Dmitri Krioukov}
\email{dima@northeastern.edu}
\affiliation{Department of Physics, Northeastern University, Boston, Massachusetts, USA 02115}
\affiliation{Department of Mathematics, Northeastern University, Boston, Massachusetts, USA 02115}
\affiliation{Department of Electrical \& Computer Engineering, Northeastern University, Boston, Massachusetts, USA 02115}
\affiliation{Network Science Institute, Northeastern University, Boston, Massachusetts, USA 02115}

\date{\today}

\begin{abstract}
Poisson sprinkling provides a Lorentz-invariant way of generating causal sets from a continuum spacetime, but numerical work necessarily restricts the sprinkling to a finite region. For observables involving causal-set links, this finite-volume restriction can generate a bias that does not disappear by simply increasing the number of sprinkled points. The origin of the problem is geometric. At fixed proper time from an event in Minkowski spacetime, candidate links populate a non-compact hyperbolic space of rapidities. In dimensions $d+1$ with $d\geq 2$, this space is boundary dominated: the fraction of volume close to the boundary of a large region does not vanish in the infinite-volume limit. Different finite-volume regularizations can therefore produce different limiting normalized statistics. We relate this phenomenon to the failure of naive open-boundary thermodynamic limits on non-amenable spaces, derive explicitly the dependence of the link proper-time distribution on the finite cutoff, and distinguish a well-defined link-intensity profile from the ill-defined notion of a uniformly chosen link. We verify these results with independent Poisson-sprinkling simulations in $d+1$ dimensions with $d=1,2,3$. We then propose an event-centered product-window estimator with a controlled double-scaling limit, and discuss compact hyperbolic quotients as a possible boundary-free alternative.
\end{abstract}

\maketitle

\section{Introduction}
\label{sec:intro}

Numerical calculations in an infinite homogeneous space usually proceed by simulating a finite region and then increasing its size. In Euclidean systems, this procedure is generally justified by the fact that sufficiently large regions are dominated by their bulk, while the relative contribution of the boundary becomes negligible. This property, however, is not shared by all geometries. To make this distinction precise, let $\{\Omega_R\}$ be an increasing family of finite regions approaching the whole non-compact space as $R\to\infty$, a construction usually referred to as an \emph{exhaustion}. In statistical mechanics, an exhaustion is said to be of van Hove type if, for every fixed distance $s>0$, the fraction of its volume contained in a boundary layer of thickness $s$ tends to zero~\cite{Ruelle1969},
\[
\frac{\operatorname{Vol}(\partial_s\Omega_R)}
{\operatorname{Vol}(\Omega_R)}
\longrightarrow 0.
\]
Here $\partial_s\Omega_R$ denotes the part of $\Omega_R$ lying within distance $s$ of its boundary. The closely related F{\o}lner condition provides the standard geometric characterization of \emph{amenability}: in the present setting, an amenable space is one that admits arbitrarily large regions whose boundary becomes negligible compared with their volume~\cite{Engel2018}. These notions provide a precise way of determining when a finite-domain calculation can be expected to reproduce the bulk properties of the corresponding infinite system.

Euclidean geometry is the simplest example. For a ball of radius $R$ in $\R^d$, the volume grows as $R^d$ whereas the area of the boundary grows as $R^{d-1}$. Consequently, a boundary layer of any fixed thickness occupies a vanishing fraction of the system as $R\to\infty$. Euclidean space is therefore amenable, and ordinary balls provide a van Hove/F{\o}lner exhaustion. Local observables measured in a sufficiently large bulk region then become insensitive to how the external boundary is implemented.

Negative curvature changes this logic qualitatively. For $d\geq2$, hyperbolic space $\Hh^d$ is non-amenable and therefore admits no F{\o}lner exhaustion~\cite{Engel2018}. As a consequence, the boundary of a large region never becomes negligible compared with its bulk. The origin of this statement is particularly transparent for geodesic balls. Consider a ball $B_R$ in $d$-dimensional hyperbolic space with curvature radius $\ell_H$. Its volume and boundary area satisfy
\begin{align}
V_d(R) &= \Omega_{d-1}\ell_H^d\int_0^{R/\ell_H}\!\sinh^{d-1}u\,\dd u,\\
A_d(R) &= \Omega_{d-1}\ell_H^{d-1}\sinh^{d-1}(R/\ell_H),
\end{align}
and, for $d>1$,
\begin{equation}
\frac{A_d(R)}{V_d(R)}\longrightarrow \frac{d-1}{\ell_H}.
\label{eq:cheeger}
\end{equation}
Thus the boundary remains extensive. Equivalently, a layer of fixed thickness $s$ adjacent to the boundary contains the asymptotic fraction
\begin{equation}
\frac{V_d(R)-V_d(R-s)}{V_d(R)}
\longrightarrow 1-e^{-(d-1)s/\ell_H},
\label{eq:boundarylayer}
\end{equation}
which is nonzero. Thus even a layer of fixed proper thickness contains a finite fraction of a large hyperbolic ball. This is the boundary-dominated behavior encoded by non-amenability and explains why an open finite patch need not represent the bulk of the infinite space even when the patch is very large. The same issue is familiar on trees: a finite Cayley tree is dominated by its leaves and is not interchangeable with the infinite Bethe lattice in the naive thermodynamic limit~\cite{Eggarter1974,LyonsPeres2016}.

These facts are not merely mathematical subtleties. Statistical-mechanical simulations on hyperbolic lattices exhibit strong dependence on boundary conditions because a finite fraction of degrees of freedom remains close to the boundary. Periodic boundary conditions on compact hyperbolic quotients provide a way to study bulk behavior without an extensive open boundary~\cite{SaussetTarjus2007}. Compact hyperbolic surfaces have also been used to avoid boundary effects in simulations of random geometric graphs~\cite{Hoorn:2021aa}, while computable distance formulas and algorithms on such surfaces provide tools for their numerical implementation~\cite{Stepanyants_2024}. Studies of the Ising model illustrate the statistical-mechanical consequences: open hyperbolic patches do not provide the same finite-size scaling construction as closed surfaces of increasing area~\cite{Breuckmann2020}.

The sampling problem considered here is closely related, but with an important twist. Minkowski spacetime itself is not a non-amenable space under translations. The difficulty appears because the set of events at fixed proper time from a given event is a hyperboloid, and the natural rapidity coordinates on that hyperboloid carry the geometry of $\Hh^d$. A Lorentz-invariant Poisson sprinkling of Minkowski spacetime therefore produces a causal set in which every event has infinitely many links in its future and past~\cite{Surya2019}. This infinite valency is a direct consequence of combining discreteness with Lorentz invariance; indeed, no intrinsic Lorentz-equivariant prescription can reduce a sprinkling to a finite-valency graph without introducing additional structure~\cite{BombelliHensonSorkin2009}.

The central question is then not simply whether a finite sprinkling is ``large enough.'' The question is how a finite sprinkling approaches the infinite-sprinkling limit. To answer this question, we have to distinguish between different types of limits. If our spacetime is of a finite fixed volume, then we can talk only about the continuum limit, in which the sprinkling density~$\rho$ grows to infinity. This limit does not cause any boundary problems. If, however, the spacetime has infinite volume, then we can approach the infinite-sprinkling limit in different ways. One way is first to fix~$\rho$, and then consider sprinklings in an exhaustive sequence of spacetime chunks whose finite volume grows to infinity so that they eventually cover the whole infinite spacetime in the limit. We call this limit infrared or infinite-volume limit. Upon reaching it, we can then tend $\rho$ to infinity to arrive at the continuum limit in the whole infinite spacetime. The other way is first to take the continuum limit in a finite-volume chunk, and then grow it to fixity. Yet another option is to start with a constant-$\rho$ sprinkling in the whole infinite spacetime, thus sprinkling is infinite from the very beginning, and then tend $\rho$ to infinity to reach the continuum limit.

The key point we aim to demonstrate in this paper is that none of these limit-taking procedures are equivalent in general. In particular, the infrared and continuum limit do not commute. We note that noncommutativity between infrared and continuum limits already appears in $(1+1)$-dimensional Minkowski spacetime for standard nonlocal causal-set d'Alembertians acting on noncompact fields~\cite{BogunaKrioukov2026}. This observation makes it important to separate generic order-of-limits effects from the specifically hyperbolic boundary mechanism that we study here. More importantly, for observables involving the infinitely many links of one event, different ways of exhausting the non-compact rapidity directions can yield different limiting normalized distributions. In this sense, the boundary problem is intrinsic to the observable rather than a conventional finite-size correction.

In a similar vein, Dowker, Liu, and Lloyd-Jones~\cite{DowkerLiuLloydJones2025} have recently provided analytical and numerical evidence that timelike boundaries produce contributions to the mean Benincasa--Dowker--Glaser causal-set action that diverge in the continuum limit. Although this is a different phenomenon from the sampling bias studied here, it similarly illustrates that increasing the sprinkling density does not necessarily suppress boundary effects.

In the following we make our boundary sampling bias statements explicit for the proper-time distribution of links. The main conclusion is that the expression obtained by formally cancelling an infinite rapidity volume is meaningful as a \emph{link-intensity profile per unit hyperbolic volume}, but not as the probability distribution of a link chosen uniformly from the full infinite set of links. Once this distinction is made, the numerical observations and the appropriate simulation strategy become much clearer.

\section{Link statistics in sprinkled Minkowski spacetime}
\label{sec:links}

Consider a Poisson sprinkling of density $\rho$ in $(d+1)$-dimensional Minkowski spacetime. We condition on the presence of an event $e_0$ and set it at the origin. The interior of its future light cone can be written in Milne coordinates,
\begin{equation}
\dd s^2=-\dd\tau^2+\tau^2\left(\dd\chi^2+\sinh^2\chi\,\dd\Omega_{d-1}^2\right),
\label{eq:milne}
\end{equation}
where $\tau$ is the proper time from $e_0$ and $\chi$ is a rapidity coordinate defined in the domain $\chi \in (-\infty,\infty)$ for $d=1$ and $\chi \in(0,\infty)$ for $d>1$. The spacetime volume element is
\begin{equation}
\dd V=\tau^d\,\dd\tau\,\dd V_{\Hh^d},\qquad
\dd V_{\Hh^d}=\sinh^{d-1}\chi\,\dd\chi\,\dd\Omega_{d-1}.
\end{equation}

Two sprinkled events at proper-time separation $\tau$ form a link if their Alexandrov interval contains no other sprinkled event. Since the volume of such interval is
\begin{equation}
V_{d+1}(\tau)=v_d\tau^{d+1},\qquad
v_d=\frac{2}{d(d+1)\Gamma(d/2)}\left(\frac{\pi}{4}\right)^{d/2},
\end{equation}
then the link probability between $e_0$ and one event at proper time $\tau$ is
\begin{equation}
P_{\rm link}(\tau)=e^{-\rho v_d\tau^{d+1}}.
\end{equation}
The expected link-intensity measure from $e_0$ is therefore
\begin{equation}
\dd\Lambda_{e_0}=
\rho\,\tau^d e^{-\rho v_d\tau^{d+1}}\,\dd\tau\,\dd V_{\Hh^d}.
\label{eq:intensity}
\end{equation}
Equation~\eqref{eq:intensity} is locally finite and Lorentz invariant, but its total mass is infinite because $\mathrm{Vol}(\Hh^d)=\infty$. This is the precise origin of the infinite valency of a sprinkled Minkowski causal set.

Let us now integrate Eq.~\eqref{eq:intensity} over rapidity, cancel the same infinite factor in numerator and denominator, and define
\begin{equation}
q_\rho(\tau)=(d+1)\rho v_d\tau^d e^{-\rho v_d\tau^{d+1}}.
\label{eq:qprofile}
\end{equation}
Introducing the dimensionless time $\hat\tau \equiv \tau(\rho v_d)^{1/(d+1)}$, Eq.~\eqref{eq:qprofile} becomes
\begin{equation}
q(\hat\tau)=(d+1)\hat\tau^d e^{-\hat\tau^{d+1}}.
\label{eq:qscaled}
\end{equation}
We now interpret this distribution. Because there are infinitely many links, there is no normalized uniform probability measure on the link set. Equation~\eqref{eq:qscaled} should instead be understood as the normalized $\tau$-profile of the intensity \emph{per unit hyperbolic volume}. Equivalently, it is the distribution obtained from any finite cutoff whose allowed hyperbolic volume is independent of $\tau$ before the cutoff is removed.

This distinction is precisely where finite simulations enter. A finite spacetime patch does not generally impose a $\tau$-independent rapidity cutoff. The infinite factor can therefore not be cancelled before specifying how the infinite region is approached.

\section{Finite regulators, exhaustions, and sampling bias}
\label{sec:finite}

Before proceeding, we need to fix some terminology. In the usual physics sense, a \emph{regulator} is a finite cutoff introduced to make an otherwise infinite or ill-defined quantity finite. Here the relevant cutoffs are infrared ones: for example, a finite upper bound for coordinate time $T$ limits the temporal extent of the sprinkling and $\chi_c$ limits the rapidity range. The finite domain obtained after imposing the cutoff will be called the \emph{regulated domain}, while the family of such domains obtained as the cutoff is removed defines an exhaustion of the corresponding infinite space. This distinction is important: the sprinkling density $\rho$ is not an infrared regulator. Increasing $\rho$ at fixed physical domain is a continuum, or large-density, limit, whereas increasing the size of the regulated domain is an infinite-volume limit.

Let a finite rapidity cutoff restrict the coordinate to $0\leq\chi\leq\chi_c(\tau)$. The measured proper-time density is then proportional to
\begin{equation}
p_{\rm reg}(\tau)\propto
\tau^d e^{-\rho v_d\tau^{d+1}}
\mathcal V_d\!\left[\chi_c(\tau)\right],
\label{eq:preggeneral}
\end{equation}
where
\begin{equation}
\mathcal V_d(\chi)=\Omega_{d-1}\int_0^\chi\sinh^{d-1}u\,\dd u
\end{equation}
is the volume of a hyperbolic ball of radius $\chi$. A product cutoff $\chi_c=\mathrm{const.}$ reproduces Eq.~\eqref{eq:qscaled}. A conventional finite-volume regularization in Minkowski coordinates does not, because it induces a proper-time-dependent rapidity cutoff.

To illustrate the last point, consider the finite region
\begin{equation}
\mathcal R_T=J^+(e_0)\cap\{t<T\},
\label{eq:cone}
\end{equation}
that is, the future cone~$J^+$ of~$e_0$ truncated by a constant-time plane $t=T$. Since $t=\tau\cosh\chi$ and $t<T$, events at proper time $\tau$ have rapidities bounded by
\begin{equation}
\chi_c(\tau)=\operatorname{arcosh}(T/\tau)
\simeq \ln\left(\frac{2T}{\tau}\right)
\qquad (T\gg\tau).
\label{eq:chicut}
\end{equation}
This particular finite-volume regularization can be treated without approximation. Defining
\begin{equation}
\hat T \equiv T(\rho v_d)^{1/(d+1)},
\end{equation}
the normalized distribution measured in $\mathcal R_T$ is
\begin{equation}
p_T(\hat\tau)=\frac{1}{Z_d(\hat T)}
\hat\tau^d e^{-\hat\tau^{d+1}}
\mathcal V_d\!\left[
\operatorname{arcosh}\left(\frac{\hat T}{\hat\tau}\right)
\right],
\label{eq:pTexact}
\end{equation}
where $Z_d(\hat T)$ is fixed by normalization. The angular factor contained in $\mathcal V_d$ therefore cancels from the normalized distribution.

For $d>1$,
\begin{equation}
\mathcal V_d[\chi_c(\tau)]\propto e^{(d-1)\chi_c}
\propto \left(\frac{T}{\tau}\right)^{d-1},
\end{equation}
so that the factor $\tau^d$ in Eq.~\eqref{eq:preggeneral} is reduced to a single power of $\tau$. In rescaled variables, the limiting distribution is
\begin{equation}
p_T(\hat\tau)=
\frac{d+1}{\Gamma\!\left(2/(d+1)\right)}
\hat\tau\,e^{-\hat\tau^{d+1}},\qquad d>1.
\label{eq:ptdgt1}
\end{equation}
The dependence on $T$ cancels after normalization, so increasing the box size does not recover Eq.~\eqref{eq:qscaled}. Removing the cutoff along this exhaustion therefore leads to a limiting distribution different from the one obtained with a $\tau$-independent rapidity cutoff. In other words, the limiting normalized observable retains information about how the infinite rapidity domain has been approached.

The case $d=1$ is qualitatively different. Here $\Hh^1$ is just a line and
\begin{equation}
p_T(\hat\tau)\propto
\hat\tau e^{-\hat\tau^2}
\ln\left(\frac{2\hat T}{\hat\tau}\right).
\label{eq:ptd1}
\end{equation}
After normalization, the $\ln(2\hat T)$ term dominates as $\hat T\to\infty$, and Eq.~\eqref{eq:ptd1} converges slowly to $2\hat\tau e^{-\hat\tau^2}$, which is precisely Eq.~\eqref{eq:qscaled} for $d=1$. Thus the persistent bias occurs for $d\geq2$, exactly the dimensions for which the rapidity space $\Hh^d$ is non-amenable. The $(1+1)$-dimensional case is qualitatively different. There, the cutoff dependence is only logarithmic and vanishes in the infinite-volume limit, whereas for $d\geq2$ the exponential growth of hyperbolic volume leaves a finite boundary contribution.

It is important to stress that this conclusion is specific to the normalized link statistics considered here. As shown in Ref.~\cite{BogunaKrioukov2026}, a different order-of-limits problem occurs already in $1+1$-dimensional Minkowski spacetime for the nonlocal causal-set d'Alembertian introduced in Ref.~\cite{Sorkin2009}. For a constant scalar field restricted to a finite spatial interval $|x|\leq L$, the expectation value converges to zero for every fixed finite $L$ as $\rho\to\infty$, whereas taking $L\to\infty$ first gives the noncompact-field result $\langle B_{\rho}^{(2)}\phi\rangle=-2\rho\phi_0$, which diverges in the continuum limit. Thus the amenability of $\Hh^1\simeq\R$ does not prevent noncommuting infrared and continuum limits. The origin there is the nonuniform convergence of a nonlocal operator with respect to its infrared regulator. The mechanism studied here is more specific: after normalization the finite-cone bias disappears in $d=1$, while for $d\geq2$ non-amenability keeps the boundary contribution extensive and the regulator dependence persists.

\section{What should be sampled?}
\label{sec:what}

The preceding calculation suggests that two questions that look similar must be kept separate.

First, if the goal is an intrinsic Lorentz-invariant statement about the sprinkled causal set, then a ``uniformly random link'' is not an admissible observable. The link set is infinite and the Lorentz group acts transitively on each constant-$\tau$ hyperboloid. There is no finite invariant measure on that orbit. The invariant object is the sigma-finite intensity measure in Eq.~\eqref{eq:intensity}, or quantities obtained from it by integrating over finite invariant spacetime regions. In this interpretation Eq.~\eqref{eq:qscaled} is a useful proper-time profile per unit hyperbolic volume, not a probability law on the full link set.

Second, if the goal is numerical estimation of that proper-time profile, then one is free to introduce a finite cutoff, but the associated regularization---equivalently, the chosen way of exhausting the infinite rapidity space---must be regarded as part of the estimator.

A natural choice is an event-centered \emph{product window}. For every root event $x$, choose a timelike direction $u_x$ defining rapidity and retain only links connecting $x$ and $y$ satisfying
\begin{equation}
0<\tau(x,y)<\tau_c,\qquad 0<\chi_{u_x}(x,y)<\chi_c.
\label{eq:localwindow}
\end{equation}
Because the rapidity volume then factorizes from the proper-time dependence, the measured $\hat\tau$ distribution is exactly
\begin{equation}
p_{\rm pw}(\hat\tau)=
\frac{(d+1)\hat\tau^d e^{-\hat\tau^{d+1}}}
{1-e^{-\hat\tau_c^{d+1}}},
\qquad 0<\hat\tau<\hat\tau_c,
\label{eq:ppw}
\end{equation}
where $\hat\tau_c \equiv (\rho v_d)^{1/(d+1)}\tau_c$. Therefore Eq.~\eqref{eq:qscaled} is recovered whenever
\begin{equation}
\rho v_d\tau_c^{d+1}\gg1.
\label{eq:tauccondition}
\end{equation}
The expected number of retained future links is
\begin{equation}
\bar k(\chi_c,\tau_c)=
\frac{\mathcal V_d(\chi_c)}{(d+1)v_d}
\left(1-e^{-\rho v_d\tau_c^{d+1}}\right),
\label{eq:kbar}
\end{equation}
so that $\bar k\sim e^{(d-1)\chi_c}$ for $d>1$, whereas $\bar k\sim\chi_c$ for $d=1$.

There is, however, an important caveat. Rapidity relative to a compact hyperbolic ball requires a timelike reference direction. A finite $\chi_c$ therefore introduces a preferred frame. This does not invalidate the method as a \emph{numerical estimator} of the proper-time intensity profile, but it should not be confused with an intrinsic Lorentz-invariant truncation of the causal set. The theorem of Bombelli, Henson, and Sorkin makes precisely this distinction sharp~\cite{BombelliHensonSorkin2009}.

\section{A controlled double-scaling prescription}
\label{sec:scaling}

We now make a specific prescription based on the considerations in the previous section. Suppose the ambient sprinkling is generated in a finite region bounded in the future by $t=T$. Let
\begin{equation}
\Delta T_x=T-t_x
\end{equation}
be the time available above a root event $x$. A local product window remains entirely inside the simulation domain provided that
\begin{equation}
\tau_c\cosh\chi_c\leq \Delta T_x .
\label{eq:fit}
\end{equation}
This geometric constraint can be reconciled with Eq.~\eqref{eq:tauccondition} by taking a double-scaling limit in which the rapidity cutoff grows, while the proper-time cutoff remains large compared with the characteristic link scale.

For definiteness, we choose
\begin{equation}
\tau_c=\eta\,\frac{\Delta T_x}{\cosh\chi_c},
\qquad 0<\eta<1,
\label{eq:taucchoice}
\end{equation}
so that the product window remains strictly inside the ambient simulation region. Since the natural macroscopic length scale of the simulation is its temporal extent $T$, we let the rapidity cutoff grow as
\begin{equation}
\chi_c(\rho,T)=\frac{\mu}{d+1}
\ln\!\left(\rho T^{d+1}\right),
\qquad 0<\mu<1.
\label{eq:chiscaling}
\end{equation}
The restriction $0<\mu<1$ follows directly from the two requirements of the construction. First, $\mu>0$ guarantees that
\begin{equation}
\chi_c\longrightarrow\infty
\end{equation}
as $\rho T^{d+1}\to\infty$, so that the rapidity cutoff is progressively removed. On the other hand, increasing $\chi_c$ decreases the largest proper-time window that can fit inside the finite simulation. For large $\chi_c$,
\begin{equation}
\cosh\chi_c
\sim \frac{1}{2}e^{\chi_c}
=
\frac{1}{2}
\left(\rho T^{d+1}\right)^{\mu/(d+1)} ,
\end{equation}
and therefore Eq.~\eqref{eq:taucchoice} gives
\begin{equation}
\tau_c
\sim
2\eta\,\Delta T_x
\left(\rho T^{d+1}\right)^{-\mu/(d+1)} .
\end{equation}
Comparing this scale with the characteristic link scale
$\rho^{-1/(d+1)}$, we obtain
\begin{equation}
\frac{\tau_c}{\rho^{-1/(d+1)}}
\sim
2\eta\,
\frac{\Delta T_x}{T}
\left(\rho T^{d+1}\right)^{\frac{1-\mu}{d+1}} .
\label{eq:scale_ratio}
\end{equation}
Thus, for roots such that $\Delta T_x/T$ remains of order unity, the proper-time cutoff becomes parametrically larger than the characteristic link scale only if $\mu<1$. The two requirements therefore lead naturally to the condition
\begin{equation}
0<\mu<1.
\end{equation}

Equivalently, raising Eq.~\eqref{eq:scale_ratio} to the power $d+1$ gives
\begin{equation}
\rho\tau_c^{d+1}
\sim
(2\eta)^{d+1}
\left(\frac{\Delta T_x}{T}\right)^{d+1}
\left(\rho T^{d+1}\right)^{1-\mu}
\longrightarrow\infty ,
\label{eq:doublescale}
\end{equation}
while at the same time $\chi_c\to\infty$. Hence the rapidity window becomes arbitrarily large, whereas the proper-time window still contains scales much larger than the characteristic link scale.

Roots too close to the upper boundary do not satisfy this asymptotic regime and should be excluded from the estimator. From Eq.~\eqref{eq:taucchoice}, the corresponding condition can be written, up to numerical factors of order unity, as
\begin{equation}
\rho(\Delta T_x)^{d+1}
e^{-(d+1)\chi_c}\gg1 .
\label{eq:rootcriterion}
\end{equation}
Using Eq.~\eqref{eq:chiscaling}, this becomes
\begin{equation}
\left(\frac{\Delta T_x}{T}\right)^{d+1}
\left(\rho T^{d+1}\right)^{1-\mu}
\gg1 ,
\label{eq:rootcriterion_scaled}
\end{equation}
which provides a direct criterion for selecting admissible root events at a given density and simulation size.

This prescription therefore provides a controlled estimator of the proper-time intensity profile from a finite sprinkling. The parameters $\mu$ and $\eta$ determine how the rapidity and proper-time windows scale with $\rho T^{d+1}$. Varying them offers a direct convergence test: once $\hat{\tau}_c\gg1$, the rescaled proper-time distribution should become insensitive to the cutoff parameters, whereas the number of retained links may continue to depend strongly on $\chi_c$.

\begin{figure*}[t]
\centering
\includegraphics[width=0.96\textwidth]{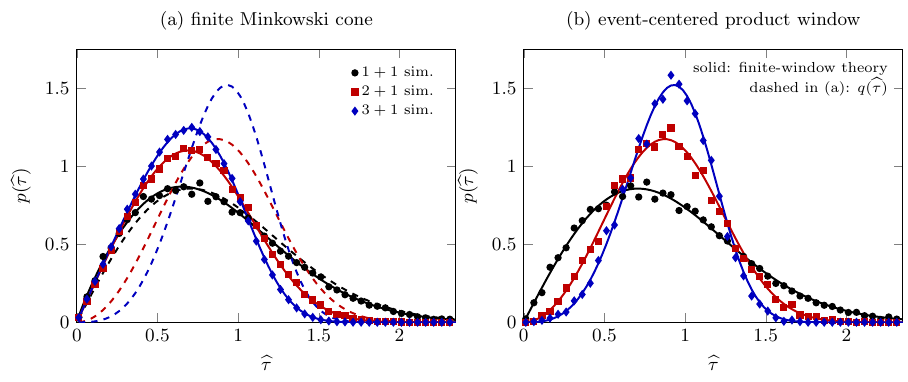}
\caption{Numerical verification of the sampling effect at $\rho=128000$ and $T=1$. Colors and symbols correspond to $1+1$ (black circles), $2+1$ (red squares), and $3+1$ (blue diamonds) dimensions. (a) Proper-time distribution of all links from the root event in the finite cone $\mathcal R_T$. Solid lines are the exact finite-cone prediction, Eq.~\eqref{eq:pTexact}, whereas dashed lines are the proper-time intensity profile $q(\hat\tau)$ in Eq.~\eqref{eq:qscaled}. (b) The same causal sets after retaining only links in the event-centered product window of Eq.~\eqref{eq:localwindow}. Solid lines are the corresponding theoretical distributions, Eq.~\eqref{eq:ppw}. For the parameters used here, the cutoff $\hat\tau_c$ lies outside the range carrying appreciable probability.}
\label{fig:simulations}
\end{figure*}

\section{Numerical simulations}
\label{sec:numerics}

We have tested the discussed prescriptions in numerical simulations. The simulations are exact in the sense that links are obtained directly from the sprinkled causal set, without using the analytical link probability as an acceptance rule.

For every realization we place a root event at the origin and sprinkle a Poisson number of events in the finite cone $\mathcal R_T$ of Eq.~\eqref{eq:cone}. Its spacetime volume is
\begin{equation}
V(\mathcal R_T)=\frac{\Omega_{d-1}}{d(d+1)}T^{d+1}.
\label{eq:conevolume}
\end{equation}
The events are sorted by Minkowski time. A candidate event $y$ is linked to the root if there is no earlier event $z$ in the Alexandrov interval between them. In Cartesian coordinates this is tested directly by requiring that no $z$ satisfies
\begin{equation}
(t_y-t_z)^2-|\bm x_y-\bm x_z|^2>0
\label{eq:linktest}
\end{equation}
together with $z\in J^+(x)\cap J^-(y)$. Since the entire interval between the root and $y$ lies inside $\mathcal R_T$, this gives the exact link relation for the sprinkled causal set.

The results shown in Fig.~\ref{fig:simulations} use $T=1$ and $\rho=128000$. For $d=1$, $2$, and $3$ we average, respectively, over $5000$, $400$, and $250$ independent sprinklings. This gives $65\,953$, $137\,512$, and $673\,778$ links from the root events. Panel (a) compares their proper-time distributions with the exact finite-cone prediction, Eq.~\eqref{eq:pTexact}. The agreement is very good in all dimensions. The proper-time intensity profile, Eq.~\eqref{eq:qscaled}, is also shown. In $2+1$ and $3+1$ dimensions the disagreement with this profile is large, as predicted by the asymptotic result in Eq.~\eqref{eq:ptdgt1}.

Panel (b) tests the product-window prescription on exactly the same sprinkled causal sets, with $\mu=0.6$ and $\eta=0.9$ in Eqs.~\eqref{eq:chiscaling} and \eqref{eq:taucchoice}. Once the proper-time and rapidity cutoffs are separated, the measured distributions collapse onto Eq.~\eqref{eq:ppw}, and therefore onto Eq.~\eqref{eq:qscaled} over the range in which the cutoff is irrelevant.

A more quantitative comparison is given in Table~\ref{tab:simulations}. The theoretical mean degree in the finite cone follows by integrating the intensity in Eq.~\eqref{eq:intensity} over $\mathcal R_T$, while the expected degree inside the product window is given by Eq.~\eqref{eq:kbar}. Both agree with the simulations at the percent level or better. More importantly, the mean proper time in the finite cone agrees with the finite-cone prediction and not with Eq.~\eqref{eq:qscaled}. In $3+1$ dimensions, for example, the simulation gives $\langle\hat\tau\rangle=0.6831$, compared with $0.6807$ from Eq.~\eqref{eq:pTexact} and $0.9064$ from Eq.~\eqref{eq:qscaled}.

\begin{table}[b]
\caption{Numerical checks at $\rho=128000$. Each entry of the last three columns is simulation/theory. The product-window parameters are $\mu=0.6$ and $\eta=0.9$.}
\label{tab:simulations}
\begin{ruledtabular}
\begin{tabular}{c r c c c}
$d+1$ & runs & $\bar k_T$ & $\langle\hat\tau\rangle_T$ & $\langle\hat\tau\rangle_{\rm pw}$\\
\hline
$1+1$ & 5000 & $10.191/10.257$ & $0.8316/0.8332$ & $0.8858/0.8862$\\
$2+1$ & 400  & $131.28/130.60$ & $0.7248/0.7290$ & $0.8930/0.8930$\\
$3+1$ & 250  & $655.11/651.42$ & $0.6831/0.6807$ & $0.9053/0.9064$\\
\end{tabular}
\end{ruledtabular}
\end{table}

These simulations provide an independent check that the effect is not produced by an approximation in the Poisson calculation. The causal sets are generated first and their links are then identified by testing interval emptiness, while Eqs.~\eqref{eq:pTexact} and \eqref{eq:ppw} are derived independently from the continuum intensity measure. The agreement between the two confirms that the discrepancy between the finite-cone distribution and Eq.~\eqref{eq:qscaled} is a sampling effect induced by the geometry of the finite-volume regularization.

\section{Discussion}
\label{sec:discussion}

The results of this paper show that finite-volume sampling of link statistics in Minkowski spacetime can lead to a bias that is qualitatively different from an ordinary finite-size correction. The origin of the effect lies in the geometry of the constant-proper-time hypersurfaces associated with a given event. These hypersurfaces are copies of hyperbolic space $\Hh^d$, and for $d\geq2$ they are non-amenable. As a consequence, the boundary of a large open region in rapidity space remains comparable with its bulk. A finite cutoff in Minkowski coordinates therefore does more than restrict the number of sampled events: it introduces a proper-time-dependent cutoff in rapidity space. After normalization, this dependence survives both the infinite-volume and large-density limits. In particular, the finite-cone distribution approaches Eq.~\eqref{eq:ptdgt1}, rather than the proper-time profile in Eq.~\eqref{eq:qscaled}.

The $(1+1)$-dimensional case is qualitatively different. There, the rapidity space is $\Hh^1\simeq\R$ and is amenable. Its volume grows only linearly with rapidity, and the dependence on the finite-cone cutoff is logarithmic. After normalization this dependence disappears as the cutoff is taken to infinity, and the distribution converges, although slowly, to Eq.~\eqref{eq:qscaled}. The persistent sampling bias found in higher dimensions is therefore directly related to the non-amenability of the rapidity space. This distinction is important: it does not imply that all problems involving the order of infinite-volume and continuum limits disappear in $(1+1)$ dimensions.

Indeed, a conceptually different order-of-limits problem occurs already in $(1+1)$-dimensional Minkowski spacetime for the nonlocal causal-set d'Alembertian introduced in Ref.~\cite{Sorkin2009}. As shown in Ref.~\cite{BogunaKrioukov2026}, when this operator acts on a constant scalar field restricted to a finite spatial interval $|x|\leq L$, taking the sprinkling density to infinity at fixed $L$ gives the continuum result,
\begin{equation}
\lim_{\rho\to\infty}
\left\langle B_{\rho,L}^{(2)}\phi\right\rangle=0.
\end{equation}
If the spatial cutoff is instead taken to infinity first, one recovers the result for a constant field over the whole noncompact spacetime,
\begin{equation}
\left\langle B_{\rho}^{(2)}\phi\right\rangle=-2\rho\phi_0,
\end{equation}
which does not converge to the continuum d'Alembertian as $\rho\to\infty$. Thus the infrared and continuum limits do not commute even though $\Hh^1$ is amenable. The origin of the effect in that case is the nonuniform convergence of a nonlocal operator with respect to its infrared cutoff. This should be distinguished from the mechanism studied here, where the observable is a normalized link distribution and the persistent dependence on the finite-volume regularization for $d\geq2$ follows from the exponential growth of hyperbolic volume. Order-of-limits problems are therefore more general than non-amenability, while non-amenability provides the specific geometric mechanism responsible for the sampling bias considered in this work.

This interpretation also clarifies what should be regarded as the intrinsic quantity in the infinite sprinkling. Since an event has infinitely many links, there is no uniform probability measure on its complete link set. Equation~\eqref{eq:qscaled} should instead be interpreted as the normalized proper-time profile of the Lorentz-invariant link intensity per unit hyperbolic volume. A finite simulation necessarily introduces an infrared cutoff, and the way in which this cutoff is taken to infinity becomes part of the sampling prescription. The numerical simulations presented here illustrate this distinction directly: sprinklings in a finite constant-time cone reproduce the finite-cone prediction in Eq.~\eqref{eq:pTexact}, whereas an event-centered product window reproduces Eq.~\eqref{eq:ppw} and approaches Eq.~\eqref{eq:qscaled} when $\widehat{\tau}_c\gg1$.

For numerical measurements of the proper-time link profile, the event-centered product window therefore provides a simple controlled prescription. By imposing independent cutoffs in proper time and rapidity, the rapidity volume factorizes from the proper-time dependence and does not distort the normalized distribution. The double-scaling construction in Eqs.~\eqref{eq:taucchoice}--\eqref{eq:doublescale} makes this prescription compatible with a finite ambient sprinkling: the rapidity cutoff can be increased while the proper-time window remains large compared with the characteristic link scale. The price is that defining rapidity with respect to a finite $\chi_c$ requires a timelike reference direction and therefore introduces a preferred frame. For the purpose considered here this is not problematic, since the construction is used as a numerical estimator rather than as an intrinsic Lorentz-invariant truncation of the causal set.

A conceptually different possibility would be to remove the spatial boundary rather than control its contribution. In hyperbolic statistical mechanics, this can be achieved by considering compact quotients of hyperbolic space, which play a role analogous to periodic boundary conditions~\cite{SaussetTarjus2007,Breuckmann2020}. A corresponding Lorentzian construction could be based on compact quotients of the hyperbolic Milne slices, together with a finite interval of proper time. Such spacetimes are locally Minkowskian but have nontrivial global topology. One could then consider a sequence of quotients whose injectivity radius grows compared with the characteristic link scale, providing a possible boundary-free route to bulk local statistics. This construction is more involved and does not preserve the full global symmetry of Minkowski spacetime, but it could provide a useful independent test of the boundary effects discussed here.

More generally, the results emphasize that specifying a finite simulation region is not merely a numerical detail when the underlying geometry is boundary dominated. The observable, the finite-volume regularization, and the order in which the infinite-volume and continuum limits are taken must all be stated explicitly. In such situations, simply increasing the number of sprinkled points or making the simulation region larger does not, by itself, guarantee convergence to the intended infinite-spacetime observable.

\begin{acknowledgments}
We thank Fay Dowker and Rafael Sorkin for useful discussions and suggestions. M.B.\ acknowledges support from grant PID2022-137505NB-C22 funded by MCIN/AEI/10.13039/501100011033. M.B.\ acknowledges the Academia Excellence award, funded by the Generalitat de Catalunya. D.K.\ acknowledges NSF Grant No.~CCF-2311160. 
\end{acknowledgments}

\bibliography{sampling_biases}

\end{document}